\pdfoutput=1
\documentclass[11pt,a4paper]{article}
\usepackage{jheppub}
\usepackage{microtype}
\usepackage{dcolumn}
\usepackage{booktabs}

\usepackage{xspace}

\def\bfseries{\fontseries \bfdefault \selectfont \boldmath}

\newcommand{\nue}{\ensuremath{\bar{\nu}_e}\xspace}

\begin{document}
 
\title{Reactor Rate Modulation oscillation analysis with two detectors in Double Chooz}


%
%
\newcommand{\Aachen}{III. Physikalisches Institut, RWTH Aachen University, 52056 Aachen, Germany}
\newcommand{\Alabama}{Department of Physics and Astronomy, University of Alabama, Tuscaloosa, Alabama 35487, USA}
\newcommand{\Argonne}{Argonne National Laboratory, Argonne, Illinois 60439, USA}
\newcommand{\APC}{APC, Universit{\'e} de Paris, CNRS, Astroparticule et Cosmologie, F-75006, Paris}
\newcommand{\CBPF}{Centro Brasileiro de Pesquisas F\'{i}sicas, Rio de Janeiro, RJ, 22290-180, Brazil}
\newcommand{\CENBG}{Universit\'e de Bordeaux, CNRS/IN2P3, CENBG, F-33175 Gradignan, France}
\newcommand{\Chicago}{The Enrico Fermi Institute, The University of Chicago, Chicago, Illinois 60637, USA}
\newcommand{\CIEMAT}{Centro de Investigaciones Energ\'{e}ticas, Medioambientales y Tecnol\'{o}gicas, CIEMAT, 28040, Madrid, Spain}
\newcommand{\Columbia}{Columbia University; New York, New York 10027, USA}
\newcommand{\Davis}{University of California, Davis, California 95616, USA}
\newcommand{\Drexel}{Department of Physics, Drexel University, Philadelphia, Pennsylvania 19104, USA}
\newcommand{\Hiroshima}{Hiroshima Institute of Technology, Hiroshima, 731-5193, Japan}
\newcommand{\IIT}{Department of Physics, Illinois Institute of Technology, Chicago, Illinois 60616, USA}
\newcommand{\INR}{Institute of Nuclear Research of the Russian Academy of Sciences, Moscow 117312, Russia}
\newcommand{\CEA}{IRFU, CEA, Universit\'{e} Paris-Saclay, 91191 Gif-sur-Yvette, France}
\newcommand{\Kansas}{Department of Physics, Kansas State University, Manhattan, Kansas 66506, USA}
\newcommand{\Kitasato}{Department of Physics, Kitasato University, Sagamihara, 252-0373, Japan}
\newcommand{\Kobe}{Department of Physics, Kobe University, Kobe, 657-8501, Japan}
\newcommand{\Kurchatov}{NRC Kurchatov Institute, 123182 Moscow, Russia}
\newcommand{\MIT}{Massachusetts Institute of Technology, Cambridge, Massachusetts 02139, USA}
\newcommand{\MaxPlanck}{Max-Planck-Institut f\"{u}r Kernphysik, 69117 Heidelberg, Germany}
\newcommand{\NotreDame}{University of Notre Dame, Notre Dame, Indiana 46556, USA}
\newcommand{\IPHC}{IPHC, CNRS/IN2P3, Universit\'{e} de Strasbourg, 67037 Strasbourg, France}
\newcommand{\SUBATECH}{SUBATECH, CNRS/IN2P3, Universit\'{e} de Nantes, IMT-Atlantique, 44307 Nantes, France}
\newcommand{\Tennessee}{Department of Physics and Astronomy, University of Tennessee, Knoxville, Tennessee 37996, USA}
\newcommand{\TohokuUni}{Research Center for Neutrino Science, Tohoku University, Sendai 980-8578, Japan}
\newcommand{\TohokuGakuin}{Tohoku Gakuin University, Sendai, 981-3193, Japan}
\newcommand{\TokyoInst}{Department of Physics, Tokyo Institute of Technology, Tokyo, 152-8551, Japan }
\newcommand{\TokyoMet}{Department of Physics, Tokyo Metropolitan University, Tokyo, 192-0397, Japan}
\newcommand{\Muenchen}{Physik Department, Technische Universit\"{a}t M\"{u}nchen, 85748 Garching, Germany}
\newcommand{\Tubingen}{Kepler Center for Astro and Particle Physics, Universit\"{a}t T\"{u}bingen, 72076 T\"{u}bingen, Germany}
\newcommand{\UFABC}{Universidade Federal do ABC, UFABC, Santo Andr\'{e}, SP, 09210-580, Brazil}
\newcommand{\UNICAMP}{Universidade Estadual de Campinas-UNICAMP, Campinas, SP, 13083-970, Brazil}
\newcommand{\vtech}{Center for Neutrino Physics, Virginia Tech, Blacksburg, Virginia 24061, USA}
\newcommand{\Chooz}{{LNCA Underground Laboratory, CNRS/IN2P3-CEA, Chooz, France}}
%
%
\newcommand{\IJC}{IJC Laboratory, CNRS/IN2P3, Universit\'e Paris-Saclay, Orsay, France}
\newcommand{\Sussex}{Department of Physics and Astronomy, University of Sussex, Falmer, Brighton, United Kingdom}
\newcommand{\Londrina}{Universidade Estadual de Londrina, 86057-970 Londrina, Brazil}
\newcommand{\Hawaii}{Physics \& Astronomy Department, University of Hawaii at Manoa, Honolulu, Hawaii, USA}
\newcommand{\IFIC}{Instituto de F\'{i}sica Corpuscular, IFIC (CSIC/UV), 46980 Paterna, Spain}
\newcommand{\KEK}{High Energy Accelerator Research Organization (KEK), Tsukuba, Ibaraki, Japan}
\newcommand{\StonyBrooks}{State University of New York at Stony Brook, Stony Brook, NY,  11755, USA}
\newcommand{\Mainz}{{Institut f\"{u}r Physik and Excellence Cluster PRISMA, Johannes Gutenberg-Universit\"{a}t Mainz, 55128 Mainz, Germany}}
\newcommand{\SK}{Kamioka Observatory, ICRR, University of Tokyo, Kamioka, Gifu 506-1205, Japan}
\newcommand{\GS}{INFN Laboratori Nazionali del Gran Sasso, 67010 Assergi (AQ), Italy}
\newcommand{\SD}{South Dakota School of Mines \& Technology, 501 E. Saint Joseph St.~Rapid City, SD 57701}
\newcommand{\Arcadia}{Physics Department, Arcadia University, 450 S. Easton Road, Glenside, PA 19038}
\newcommand{\TokyoScience}{Tokyo University of Science, Noda, Chiba, Japan}
\newcommand{\LAPP}{LAPP, CNRS/IN2P3 , 74940 Annecy-le-Vieux, France}
\newcommand{\Deceased}{Deceased.}
%
%
%
\affiliation[a]{\Aachen} 
\affiliation[b]{\Alabama} 
\affiliation[c]{\Argonne} 
\affiliation[d]{\APC} 
\affiliation[e]{\CBPF} 
\affiliation[f]{\CENBG} 
\affiliation[g]{\Chicago} 
\affiliation[h]{\CIEMAT} 
\affiliation[i]{\Drexel} 
\affiliation[j]{\INR} 
\affiliation[k]{\CEA} 
\affiliation[l]{\Kitasato} 
\affiliation[m]{\Kobe} 
\affiliation[n]{\Kurchatov} 
\affiliation[o]{\MaxPlanck} 
\affiliation[p]{\NotreDame} 
\affiliation[q]{\IPHC} 
\affiliation[r]{\SUBATECH} 
\affiliation[s]{\TohokuUni} 
\affiliation[t]{\TokyoInst} 
\affiliation[u]{\TokyoMet} 
\affiliation[v]{\Muenchen} 
\affiliation[w]{\Tubingen} 
\affiliation[y]{\UNICAMP} 
\affiliation[z]{\vtech} 
\affiliation[aa]{\Chooz} 
%
%
%
\author{The Double Chooz Collaboration:}
\author[e,d]{T.~Abrah\~{a}o,}
\author[o]{H.~Almazan,} 
\author[e]{J.C.~dos Anjos,} 
\author[v]{S.~Appel,} 
\author[k]{{J.C.~Barriere},} 
\author[a]{I.~Bekman,} 
\author[r,1]{T.J.C.~Bezerra,}\note{Now at \Sussex} 
\author[j]{L.~Bezrukov,} 
\author[g]{E.~Blucher,} 
\author[q]{T.~Brugi\`{e}re,} 
\author[o]{C.~Buck,} 
\author[b]{J.~Busenitz,} 
\author[d,aa,2]{A.~Cabrera,}\note{Now at \IJC} 
\author[h]{M.~Cerrada,} 
\author[f]{E.~Chauveau,} 
\author[e,3]{P.~Chimenti,}\note{Now at \Londrina} 
\author[k]{{O.~Corpace},} 
\author[d]{J.V.~Dawson,} 
\author[c]{Z.~Djurcic,} 
\author[n]{A.~Etenko,} 
\author[s]{H.~Furuta,} 
\author[h]{I.~Gil-Botella,} 
\author[d]{{A.~Givaudan},} 
\author[d,k]{{H.~Gomez},} 
\author[y]{L.F.G.~Gonzalez,} 
\author[c]{M.C.~Goodman,} 
\author[m]{T.~Hara,} 
\author[o]{{J.~Haser},} 
\author[a]{D.~Hellwig,} 
\author[d,4]{A.~Hourlier,}\note{Now at \MIT} 
\author[t,5]{M.~Ishitsuka,}\note{Now at \TokyoScience} 
\author[w]{J.~Jochum,} 
\author[f]{C.~Jollet,} 
\author[f,q]{K.~Kale,} 
\author[t]{M.~Kaneda,} 
\author[d]{{M.~Karakac},} 
\author[l]{T.~Kawasaki,} 
\author[y]{E.~Kemp,} 
\author[d,6]{H.~de~Kerret,}\note{\Deceased} 
\author[d]{D.~Kryn,} 
\author[t]{M.~Kuze,} 
\author[w]{T.~Lachenmaier,} 
\author[i]{C.E.~Lane,} 
\author[k,d]{T.~Lasserre,} 
\author[h]{C.~Lastoria,} 
\author[k]{D.~Lhuillier,} 
\author[e]{H.P.~Lima Jr,} 
\author[o]{M.~Lindner,} 
\author[h,13]{J.M.~L\'opez-Casta\~no,} 
\author[p]{J.M.~LoSecco,} 
\author[j]{B.~Lubsandorzhiev,} 
\author[u,m]{J.~Maeda,} 
\author[z]{C.~Mariani,} 
\author[i,7]{J.~Maricic,}\note{Now at \Hawaii} 
\author[r]{J.~Martino,} 
\author[u,8]{T.~Matsubara,}\note{Now at \KEK} 
\author[k]{G.~Mention,} 
\author[f]{A.~Meregaglia,} 
\author[i,9]{T.~Miletic,}\note{Now at \Arcadia} 
\author[i,7]{R.~Milincic,} 
\author[k,10]{A.~Minotti,} \note{Now at \LAPP} 
\author[d,h,2]{D.~Navas-Nicol\'as,}
\author[h,11]{P.~Novella,}\note{Now at \IFIC} 
\author[v]{L.~Oberauer,} 
\author[d]{M.~Obolensky,} 
\author[k]{A.~Onillon,} 
\author[n]{A.~Oralbaev,} 
\author[h]{C.~Palomares,} 
\author[e]{I.M.~Pepe,} 
\author[r,12]{G.~Pronost,}\note{Now at \SK} 
\author[b,13]{J.~Reichenbacher,}\note{Now at \SD} 
\author[o,7]{B.~Reinhold,} 
\author[v]{S.~Sch\"{o}nert,} 
\author[o]{S.~Schoppmann,} 
\author[k]{{L.~Scola},} 
\author[t]{R.~Sharankova,} 
\author[k,4]{V.~Sibille,} 
\author[j]{V.~Sinev,} 
\author[n]{M.~Skorokhvatov,} 
\author[a]{P.~Soldin,} 
\author[a]{A.~Stahl,} 
\author[b]{I.~Stancu,} 
\author[w]{L.F.F.~Stokes,} 
\author[s,d]{F.~Suekane,} 
\author[n]{S.~Sukhotin,} 
\author[u]{T.~Sumiyoshi,} 
\author[b,7]{Y.~Sun,} 
\author[k]{{C.~Veyssiere},} 
\author[r]{B.~Viaud,} 
\author[k]{M.~Vivier,} 
\author[d,e]{S.~Wagner,} 
\author[a]{C.~Wiebusch,} 
\author[c,14]{G.~Yang,}\note{Now at \StonyBrooks} 
\author[r]{F.~Yermia} 
%
%
\emailAdd{navas@lal.in2p3.fr, pau.novella@ific.uv.es}

\abstract{A $\theta_{13}$ oscillation analysis based on the observed antineutrino rates at the Double Chooz far and near detectors for different reactor power conditions is presented. This approach provides a so far unique simultaneous determination of $\theta_{13}$ and the total background rates without relying on any assumptions on the specific background contributions. The analysis comprises 865 days of data collected in both detectors with at least one reactor in operation. The oscillation results are enhanced by the use of 24.06 days (12.74 days) of reactor-off data in the far (near) detector. The analysis considers the \nue interactions up to a visible energy of 8.5 MeV, using the events at higher energies to build a cosmogenic background model considering fast-neutrons interactions and $^{9}$Li decays. The background-model-independent determination of the mixing angle yields sin$^2(2\theta_{13})=0.094\pm0.017$, being the best-fit total background rates fully consistent with the cosmogenic background model. A second oscillation analysis is also performed constraining the total background rates to the cosmogenic background estimates. While the central value is not significantly modified due to the consistency between the reactor-off data and the background estimates, the addition of the background model reduces the uncertainty on $\theta_{13}$ to 0.015. Along with the oscillation results, the normalization of the anti-neutrino rate is measured with a precision of 0.86\%, reducing the 1.43\% uncertainty associated to the expectation.} 

\maketitle
       
\clearpage

\section{Introduction}
\label{sec:intro}

In the last few decades, neutrinos have been proven to be massive particles in several oscillation experiments \cite{pdg}. The oscillations among the three active neutrino species are now well established, connecting the mass eigenstates ($\nu_1$,$\nu_2$,$\nu_3$) with the flavor eigenstates ($\nu_e$,$\nu_{\mu}$,$\nu_{\tau}$). The 3-flavor neutrino oscillations are described by means of three mixing angles ($\theta_{12}$, $\theta_{23}$, $\theta_{13}$), two independent mass square differences ($\Delta m^2_{21}$, $\Delta m^2_{31}$), and one phase responsible for the $CP$-violation in the leptonic sector ($\delta_{CP}$). After the observation of the dominant oscillations in the so-called solar \cite{sno,kamland} and atmospheric sectors \cite{sk,k2k}, respectively driven by $(\theta_{12},\Delta m^2_{21})$ and $(\theta_{23},\Delta m^2_{31})$, reactor neutrino experiments have recently observed the oscillation induced by the last mixing angle, $\theta_{13}$. Double Chooz, Daya Bay and RENO have provided precise measurements of $\theta_{13}$ \cite{dciii, db,reno,dciv}, relying on the observation of the disappearance of electron antineutrinos (\nue) generated in nuclear reactors at typical flight distances of 1-2 km. The $\theta_{13}$ value offered by reactor experiments is used as an external constraint in current and future accelerator-based experiments aiming at the measurement of $\delta_{CP}$ (see for instance \cite{t2kn}). As a consequence, reactor neutrino experiments play a major role in the search for the leptonic CP-violation.

Double Chooz and other reactor experiments detect the electron antineutrinos via the $\bar{\nu}_e\,p\rightarrow e^{+}\,n$ interaction, usually referred to as inverse beta decay (IBD). The time and spatial coincidence of the prompt positron and the delayed neutron capture signals yields a large signal-to-background ratio. However, accidental and correlated events induced by fast neutrons and cosmogenic radio-nuclides can mimic the characteristic IBD signature, becoming non-negligible backgrounds. The oscillation analyses presented in \cite{dciii,db,reno,dciv} are based on background models built assuming a number of background sources. The rate and energy spectrum of each background contribution is estimated from the data collected during reactor-on periods, and incorporated to the total background expectation. Accounting for these background models, $\theta_{13}$ is derived from the observed energy-dependent deficit of \nue with respect to a MC-based null-oscillation expectation or the unoscillated flux measured at a near detector. As a consequence, the oscillation analyses are background-model-dependent and the uncertainty on the background expectations may have a non-negligible impact on the uncertainty of $\theta_{13}$.

In this paper, an alternative background-model-independent oscillation analysis is presented: the Reactor Rate Modulation (RRM). In this approach, the comparison of the observed rate of \nue candidates with respect to the expected \nue rate in absence of oscillations is performed for different reactor operation conditions ranging from zero to full thermal power. This allows for a simultaneous determination of both $\theta_{13}$ and the total background rate, without making any consideration about the individual background sources. This technique is particularly competitive in the Double Chooz experiment, as it collects data from only two nuclear cores. In addition, a background-model-dependent result on $\theta_{13}$ is also obtained following the same RRM procedure. In this case, the precision on $\theta_{13}$ is improved by incorporating to the analysis a background model based on \cite{dciv}, providing also a consistency test for the model itself. This paper extrapolates the RRM analysis described in \cite{rrm}, which uses only the far detector of the Double Chooz experiment, to a multi-detector setup considering both the near and far detectors. The oscillation analysis also incorporates for the first time the reactor-off data collected by both detectors during 2017, which offer a constraint to the BG rate and serves as independent validation of the BG model. Beyond the neutrino oscillation, the analysis also yields a measurement of the observed rate of IBD interactions. The value is found to be fully consistent with the antineutrino flux normalization provided by Bugey-4 \cite{b4}, reducing its associated uncertainty.

This paper is organized as follows. Sec.~\ref{sec:selection} describes the selection of the \nue candidates and the corresponding expected backgrounds, while Sec.~\ref{sec:data} describes the reactor-on and reactor-off data samples used for the oscillation analysis. Sec.~\ref{sec:rrm} follows with the definition of the RRM approach and the systematic uncertainties involved. Finally, the oscillation analysis results are presented in Sec.~\ref{sec:results} with and without the background model constraint, and Sec.~\ref{sec:conclusions} concludes with an overview.

\section{\nue selection and expected backgrounds}
\label{sec:selection}

The setup of the Double Chooz experiment consists of two identical detectors measuring the \nue flux generated at the two reactors (B1 and B2, with a thermal power of 4.25 GW each) of the Chooz B  nuclear power plant, operated by \'Electricit\'e de France. The average distances between the far (FD) and near (ND) detectors and the reactor cores are $L\sim400$ m and $L\sim1050$ m, respectively. Both detectors are identical and yield effectively identical responses after calibration, thus leading to a major reduction of the correlated systematic uncertainties in the oscillation analyses. The detectors consist of a set of concentric cylinders and an outer muon veto on the top. The innermost volume ({\em neutrino target} or NT) contains 10 m$^3$ of Gd-loaded (0.1\%) liquid scintillator inside a transparent acrylic vessel. This volume is surrounded by another acrylic vessel filled with 23 m$^3$ of Gd-unloaded scintillator ({\em gamma-catcher} or GC). This second volume was originally meant to fully contain the energy deposition of gamma rays from the neutron capture on Gd and the positron annihilation in the target region. The GC is in turn contained within a third volume ({\em buffer}) made of stainless steel and filled with non-scintillating mineral oil. The surface of the buffer is covered with an array of 390 low background 10-inch PMTs. The NT, GC and buffer tank define the {\em inner detector} (ID). The ID is surrounded by the {\em inner muon veto} (IV), a 50 cm thick liquid scintillator volume equipped with 78 8-inch PMTs. Finally, the upper part of the detectors is covered by an outer muon veto (OV), made of plastic scintillator strips grouped in different modules. While the ID is meant to detect the IBD interactions and to allow for the event vertex and energy reconstruction, the IV and OV are devoted to the suppression and rejection of backgrounds.

The IBD candidates selection in the RRM analysis follows the lines described in \cite{dciv}. The selection relies on the twofold-coincidence signature of the IBD process, providing a prompt trigger ($e^+$) and a delayed trigger (neutron capture). The energy of the prompt signal (\emph{visible energy}) is directly related to the energy  of the interacting \nue: E$_{\nue}\approx$ E$_{e^{+}}$ + 0.78 MeV. While the Double Chooz detectors were originally designed to exploit the large neutron capture cross-section in Gd and the characteristic de-excitation gammas ($\sim$8 MeV), the so-called Total Neutron Capture (TnC) selection approach accounts for neutron captures in H (see \cite{dchi,dchii}), C and Gd. This implies that IBD detection volume considers both the NT and the GC, boosting the statistical sample of the \nue candidates by almost a factor of 3. After a series of muon-induced background vetoes based on the ID, IV and OV data (see \cite{dciv} for details), a time and spatial correlation between the prompt and delayed signals is required. The time difference between the signals is comprised between 0.5 us and 800 us, while the distance between the vertexes is imposed to be below 120 cm. In addition, an artificial neural network (ANN) has been developed relying on the prompt-delayed correlation to set a cut reducing the rate of random coincidences, especially in the n-H events (see \cite{dchii} for details). The energy windows considered for the prompt and delayed signals are 1.0-8.5 MeV and 1.3-10.0 MeV, respectively. Unlike in the IBD selection presented in \cite{dciv}, where the prompt energy signal is extended to 20 MeV to better constraint the background shapes, in this analysis it is restricted to be below 8.5 MeV ($>99.96\%$ of the reactor \nue). As discussed below, the IBD candidates above this energy are used to infer a background expectation. 

\begin{figure}
  \begin{center}
    \includegraphics[scale=0.37]{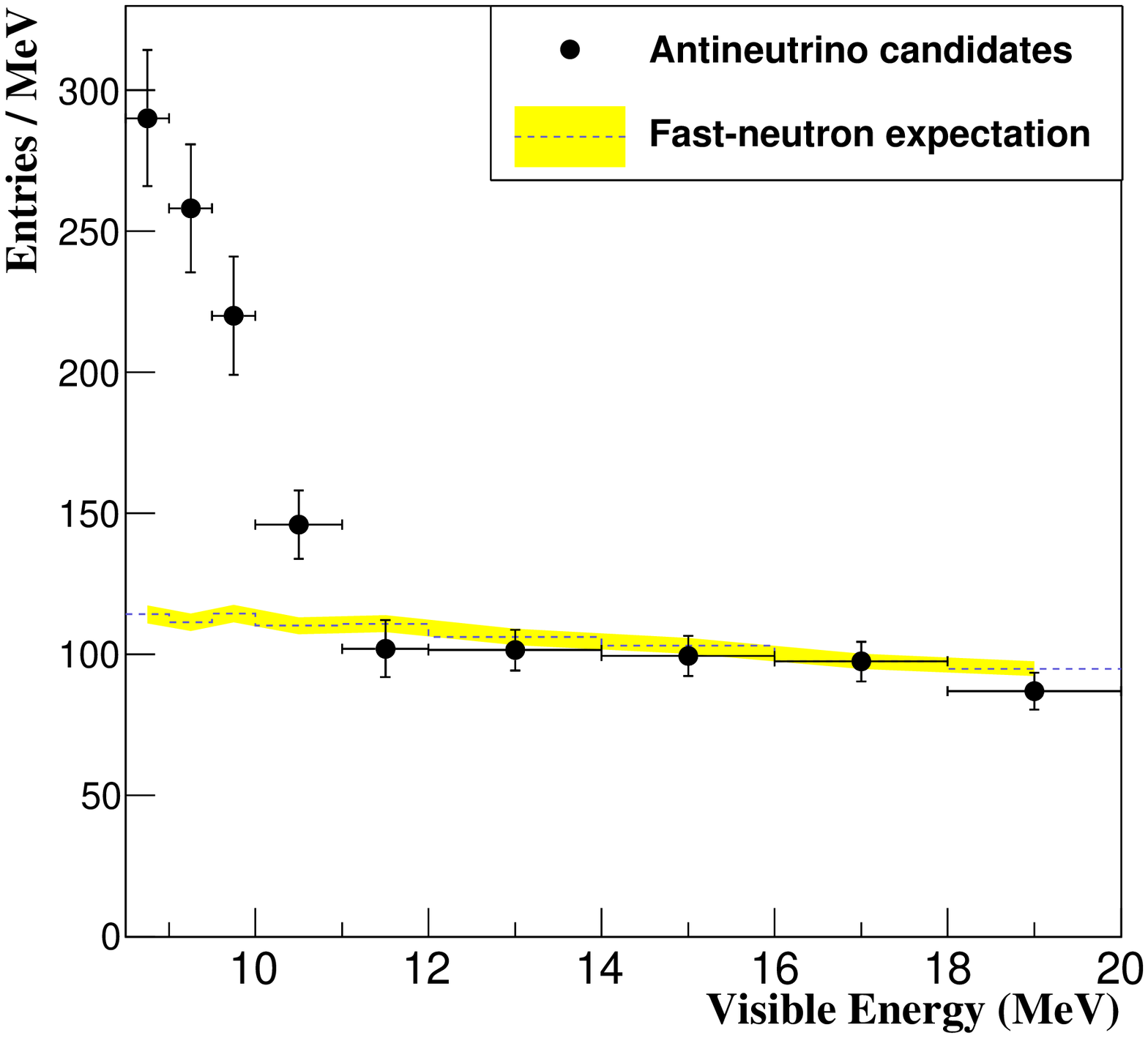}
    \includegraphics[scale=0.37]{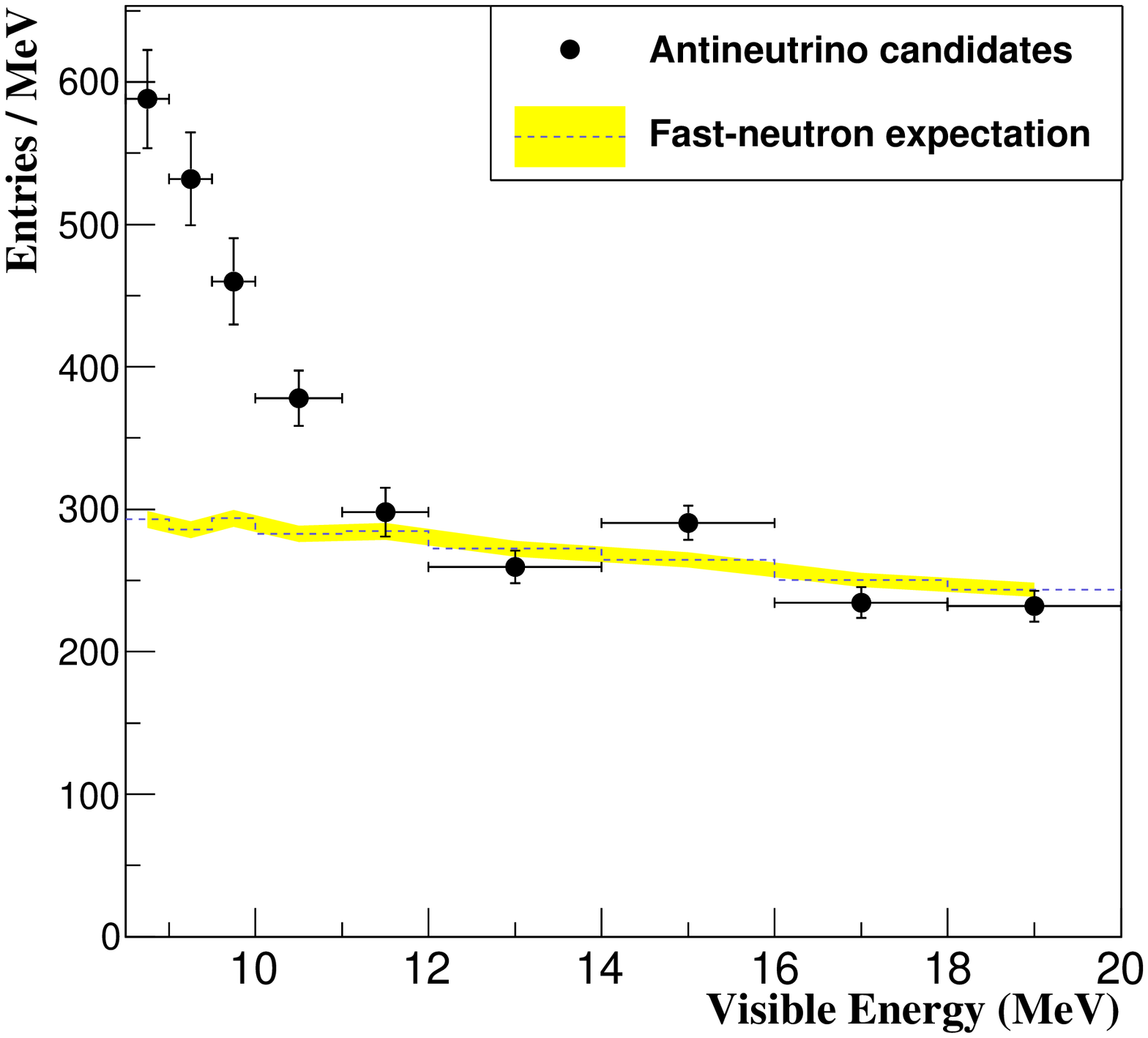}
    \caption{ IBD candidates (black dots) as a function of the visible energy above 8.5 MeV, extracted from reactor-on data in the FD (left) and the ND (right). The expected contribution from fast-neutrons is shown with dashed line and yellow error band. The fast-neutron subtraction in the 8.5-12 MeV range provides the number of $^{9}$Li decays which is used to extrapolate the total rate below 8.5 MeV.}
    \label{fig:li9}
  \end{center}
\end{figure}

The physical events mimicking the IBD signature have been discussed in \cite{dciv}. In Double Chooz, given the small overburden of the detectors (depths of $\sim$100 m and $\sim$30 m for the FD and ND, respectively), the muon-induced cosmogenic backgrounds dominate. These correspond mostly to fast-neutrons and unstable isotopes produced upon $^{12}$C spallation (mainly $^{9}$Li, as no indication of $^{8}$He is reported in \cite{li9}). While the fast-neutron twofold signature is due to a proton recoil on H followed by the n-capture, $^{9}$Li undergoes a $\beta$-n decay. Other cosmogenic backgrounds (like $\mu$ decay at rest or $^{12}$B) are estimated to be negligible. Apart from the correlated backgrounds, random coincidences of natural radioactivity events and neutron captures (hereafter, accidental background) also become a non-negligible contamination in the IBD candidates samples. However, the estimation of the accidental background contribution relying on the rate of single events is very precise ($<$0.5\% in the FD and 0.1\% in the ND).     
The reactor-on-based background model adopted in the current analysis is built with the contributions of fast-neutrons, $^{9}$Li and accidental events. The only difference with respect to the model in \cite{dciv} is the range of the visible energy and the estimation of the $^{9}$Li contribution from the candidates observed in the 8.5 MeV to 12.0 MeV energy window. The fast-neutron rate and energy spectrum is measured from events tagged by the IV up to 20 MeV. Subtracting the fast-neutron contribution in the 8.5-12.0 MeV range, as shown in Fig.~\ref{fig:li9}, offers a direct measurement of the number of $^{9}$Li decays (the \nue contribution is $0.030\pm0.009$\%). Given that the spectral shape of the $^{9}$Li prompt signal is well known \cite{li9}, the fraction of the spectrum below (above) 8.5 MeV is computed to be $89.3\pm0.5$\% ($10.7\pm0.5$\%). This number allows to extrapolate the total number of $^{9}$Li decays in the 1.0-8.5 MeV energy window considered for the RRM oscillation analysis. The expected rates for the fast-neutron, $^{9}$Li and accidental backgrounds are summarized in Tab.~\ref{tab:bg}. The background estimates are quoted separately for the first phase of the experiment (single-detector, hereafter SD), operating only the FD, and the second phase (multi-detector, hereafter MD) with both detectors running. The increase in the FD accidental rate between both periods is due to the increase of the light noise background described in \cite{lightnoise}. This noise has been suppressed in the ND covering the PMT bases with a radioupure polyester film, yielding a reduction in the accidental rate with respect to the FD. The increase in the FD accidental rate uncertainty is due to the smaller statistical sample of random coincidences used to estimate this background in the MD phase.

\begin{table}[!htb]
  \caption{\label{tab:bg} Background expectation in the 1.0-8.5 MeV window. The accidental, fast-neutron and $^{9}$Li decay contributions to the background model are derived from reactor-on IBD candidates.}
\begin{center}
\begin{tabular}{cccc}
  \hline
  Rate (day$^{-1}$) & FD (SD) & FD (MD)  & ND (MD) \\ \hline
  Accidental     & $3.930\pm0.010$ & $4.320\pm0.020$ & $3.110\pm0.004$ \\ 
  Fast-neutron   & \multicolumn{2}{c}{$1.09\pm0.03$}  & $8.89\pm0.18$ \\ 
  $^{9}$Li isotope & \multicolumn{2}{c}{$2.30\pm0.30$} & $14.09\pm1.62$ \\\hline
\end{tabular}
\end{center}
\end{table}

\section{Reactor-on and Reactor-off data samples}
\label{sec:data}

\begin{figure}
  \begin{center}
    \includegraphics[scale=0.77]{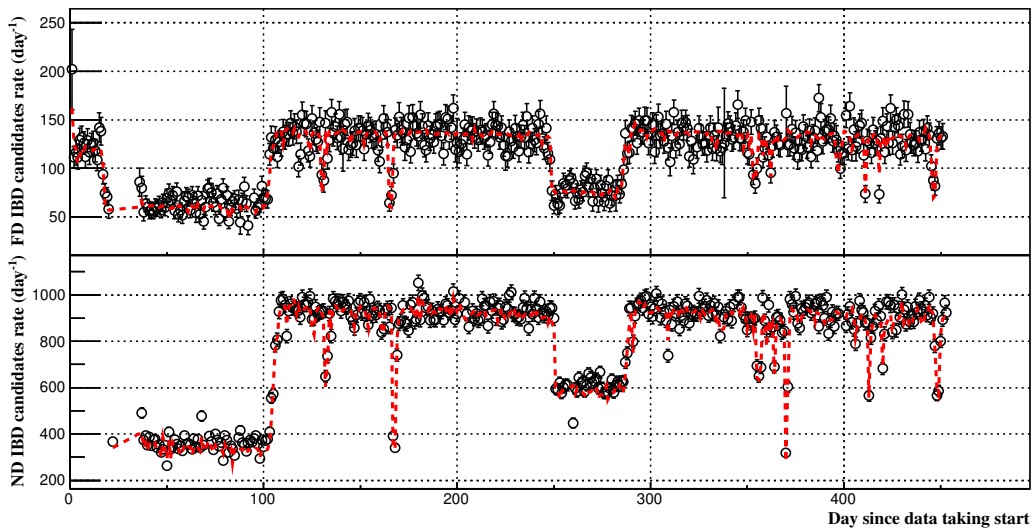}
    \caption{ IBD candidates rate for FD (top) and ND (bottom) as a function of the day of data taking during the MD period. Empty circles show the observed rate of candidates, while the dashed red line shows the expected unoscillated \nue rate according to the reactor-on simulation.  The significant rate difference in the various 1-Off periods for each detector corresponds to the different baselines between the detector and the two reactors.}
    \label{fig:nurate}
  \end{center}
\end{figure}

The Double Chooz data have been taken under different reactor operating conditions. In particular, the total \nue flux changes significantly during the reactor refuelling periods, when only one of the cores is in operation. In addition, the flux depends on the cores fuel composition, thus evolving in time. As done in \cite{dciv}, the oscillation analysis presented in this work comprises the data taken between April 2011 and January 2013 (481 days), when only the FD was available, and between January 2015 and April 2016 (384 days), when the FD and the ND were simultaneously collecting data. According to the selection described in Sec.~\ref{sec:selection}, the number of \nue candidates (actual \nue plus background contributions) in the single-detector period is 47351, while in the multi-detector period is respectively 42054 in the FD and 206981 in the ND. The time evolution of the candidates rate for the MD data samples is shown in Fig.~\ref{fig:nurate}, where the days with one (1-Off) and two (2-On) reactors in operation are clearly visible. A prediction of the unoscillated \nue flux during the reactor-on periods has been carried out as described in previous Double Chooz publications \cite{dciii}. The reactor flux model is adopted from \cite{huber,mueller}. While the $^{235}$U, $^{239}$Pu, and $^{241}$Pu isotopes contributions are derived from the Institut Laue-Langevin (ILL) reactor data (see for instance \cite{Schreckenbach:1985ep}), the contribution from $^{238}$U is predicted from \cite{haag}. The time evolution of the fission fractions are accounted for using dedicated Chooz reactor simulations. As done in past Double Chooz publications where the ND was not available, Bugey4 \cite{b4} data have been used as a virtual near detector to define the absolute flux normalization in SD. Although this is not required in oscillation analyses with MD data, the associated flux simulation still accounts for the Bugey4-based normalization in order to keep the consistency between the SD and MD flux expectations. This choice does not impact the oscillation analysis precision as the Bugey-4 uncertainty ($1.4\%$) is fully correlated between the FD and the ND, and therefore suppressed to a negligible level in a multi-detector analysis.  

Among the $\theta_{13}$ reactor-based oscillation experiments, Double  Chooz is unique in obtaining reactor-off data (2-Off) when the two cores of the Chooz site are brought down for refuelling or maintenance. Since the Daya Bay and RENO experiments are exposed to the \nue fluxes from 6 different cores, they have been so far unable to collect data when all of them are off. Double Chooz has taken reactor-off data samples during both the SD (two samples in 2011 and 2012) and MD (two samples in 2017) periods. The corresponding livetimes and number of IBD candidates in each period and detector are listed in Tab.~\ref{tab:2off}. In order to reduce the cosmogenic backgrounds, a time veto of 1.25 $\mu$s is applied after each detected muon. Thus, the corresponding livetimes differ in the near and far detectors during the MD period due to the different overburdens and muon rates in the experimental sites. Applying the selection cuts to the reactor-off data provides an inclusive sample of the different backgrounds, regardless of their origin, with a contribution from residual antineutrinos. After the nuclear reactors are turned off, $\beta$ decays from fission products keep taking place generating a residual flux of \nue which vanishes with time. Since the reactor-off periods are short in time, the contribution from the residual \nue flux is small but not negligible. However, the amount of residual \nue can be estimated either by means of Monte Carlo simulations, or by performing a relative comparison of the rates observed at different baselines. Thereby once the residual neutrinos are estimated, the reactor-off data allow for a direct measurement of the total background remaining in the \nue candidates samples.

\begin{table}[!htb]
\caption{\label{tab:2off} Reactor-off data samples during SD and MD. The last row shows the total background expectation between 1.0 and 8.5 MeV according to the model described in Sec.~\ref{sec:selection}, without considering the residual \nue.}
\begin{center}
\begin{tabular}{cccc}
  \hline
  Detector (period)               & FD (SD) & FD (MD) & ND (MD)\\ \hline
  Live time (day)                 & 7.16      & 16.90      & 12.74 \\
  IBD Candidates (day$^{-1}$)      & $7.96\pm1.05$ &  $7.99\pm0.69$ & $29.91\pm1.53$\\
  \hline
  Expected background (day$^{-1}$) & $7.32\pm0.30$ & $7.71\pm0.30$ & $26.09\pm1.63$\\
  \hline
\end{tabular}
\end{center}
\end{table}

The SD reactor-off data has been used in \cite{2off} to estimate the total background rate in previous Double Chooz IBD selection procedures, as well as to confront it with the corresponding background models. In this analysis, the data have been reprocessed with the current IBD selection, yielding the rate of candidates quoted in Tab.~\ref{tab:2off}. In order to estimate the residual neutrino contribution, the Monte-Carlo approach described in \cite{2off} has been adopted. A dedicated  simulation  has  been performed with FISPACT \cite{fispact}, an evolution code predicting the isotope inventory in the reactor cores. The neutrino spectrum is then computed using the BESTIOLE \cite{mueller} database. The expected rate of residual \nue in SD reactor-off period is found to be $0.58\pm0.18$ day$^{-1}$. Once subtracted to the observed rate of events, the measured inclusive background rate is $7.38\pm1.07$ day$^{-1}$, in good agreement with the expectation from the background model defined in Sec.~\ref{sec:selection}.

\begin{figure} 
  \begin{center}
    \includegraphics[scale=0.37]{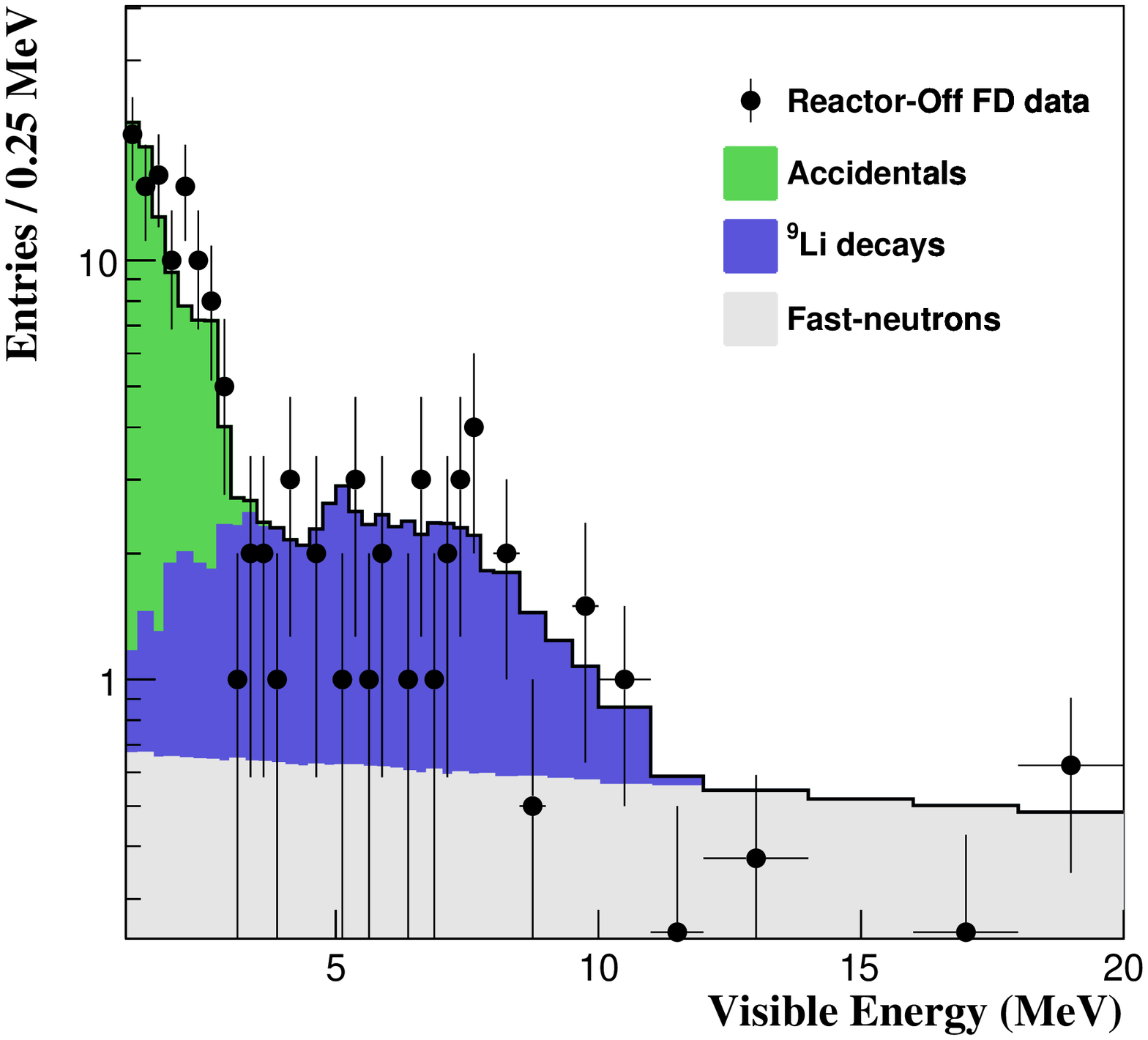}
    \includegraphics[scale=0.37]{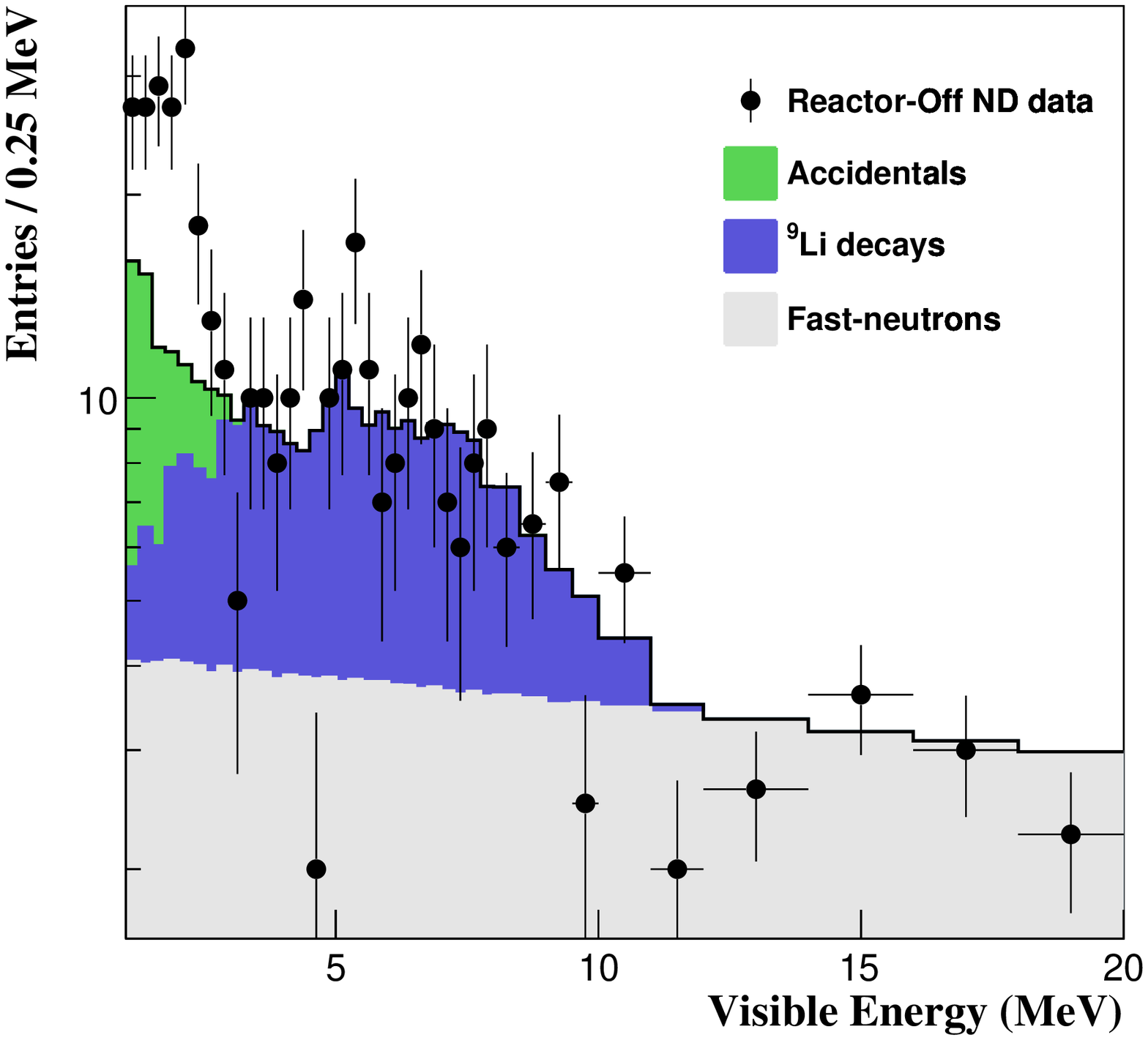}
    \caption{ Reactor-off IBD candidates as a function of the visible energy during the MD period. The observed candidates in the FD (left) and the ND (right) are superimposed to the background model described in \cite{dciv}, considering accidental coincidences, fast-neutrons and $^{9}$Li decays.}
    \label{fig:nuoff}
  \end{center}
\end{figure}

The larger reactor-off statistical sample in the MD period, especially in the ND, allows for detailed comparisons with the background model adopted in this work and described in \cite{dciv}. The energy spectra of the reactor-off IBD candidates in the FD and ND are shown in Fig.~\ref{fig:nuoff}, superimposed to the background model. The expectation reproduces the data at high energies, but deviates below $\sim$3.0 MeV. This discrepancy corresponds to the presence of the residual neutrinos, whose spectrum is known to vanish above 3 MeV due to the involved parent isotopes. According to the different geometrical acceptance between the detectors ($L_{\rm{FD}}^2/L_{\rm{ND}}^2\sim7$), the contribution of the residual \nue in the ND is significantly larger than in the FD. The observed reactor-off candidates above 8.5 MeV have been used to perform $^{9}$Li estimations following the analysis described in Sec.~\ref{sec:selection}, being the only difference that no IBD interactions are expected. The obtained  $^{9}$Li estimates up to 8.5 MeV are ($0.52\pm1.49$) day$^{-1}$ in the FD and ($9.33\pm5.28$) day$^{-1}$ in the ND, consistently with the reactor-on expectations quoted in Tab.~\ref{tab:bg}.

\section{Reactor Rate Modulation analysis}
\label{sec:rrm}

The measurement of the mixing angle $\theta_{13}$ in reactor experiments relies on the comparison of the observed prompt energy spectrum with the expectation in absence of oscillation. Both the deficit in the observed number of candidates and the distortion of the energy spectrum are accounted for in a \emph{rate plus shape} (R+S) analysis, as done in \cite{dciv}. The null-oscillation  expectation comes either from a reactor flux simulation or from the measurement of the flux at a short baseline (near detector), where the oscillation effect is small or still negligible. Besides the \nue flux, a R+S background model in each detector is also considered, assuming a certain number of background sources. Thus, the $\theta_{13}$ determination becomes background-model-dependent and the associated uncertainty contains a non-negligible contribution from the background model.

As an alternative approach, the RRM analysis offers a background-model-independent measurement of $\theta_{13}$. A \emph{rate-only} (RO) determination of the mixing angle is implemented by comparing the observed rate of candidates ($R^{\rm{obs}}$) with the expected one ($R^{\rm{exp}}$) for different reactor thermal power ($P_{ \rm{th}}$) conditions. As the signal-to-background ratio varies depending on $P_{\rm{th}}$ (the \nue flux increases with the thermal power while the background rate remains constant), the RRM analysis provides a simultaneous estimation of $\theta_{13}$ and the total (inclusive) background rate, independently of the number of background sources. Relying on rate-only information, these estimations are not affected by the energy distortion (with respect to the predicted spectrum) first reported in \cite{dciii} and further discussed in \cite{structure}. The simple experimental setup of Double Chooz, collecting 2-On, 1-Off and 2-Off data, boosts the precision of the RRM results by offering a powerful lever arm to constrain the total background. The current analysis implements for the first time a multi-detector RRM technique, using reactor-on and reactor-off data collected with both the FD and the ND. As in R+S analyses, the systematic uncertainties are highly suppressed by means of the relative comparison of the observed flux at different detectors. Beyond the error suppression, the comparison of the measurements at both detectors also provides a determination of the residual neutrino rate ($R^{\rm{r}\nu}$) in the MD reactor-off period. This ensures a precise determination of the total background rate which does not rely on Monte-Carlo simulations for the $R^{\rm{r}\nu}$ estimation.   

The RRM technique exploits the correlation of the expected and observed rates, which follows a linear model parametrized by $\sin^2(2\theta_{13})$ and the total background rate, $B_d$:

\begin{equation}
\label{eq:model} 
R_d^{\rm{obs}}=B_d+R_d^{\rm{exp}}=B_d+\left(1-\sin^2(2\theta_{13})\eta_{d}^{\rm{osc}}\right)R_d^{\rm{\nu}},
\end{equation}

\noindent where the subindex $d$ stands for either the FD or the ND, $R^{\rm{\nu}}$ is the expected rate of antineutrinos in absence of oscillation, and $\eta^{\rm{osc}}$ is the average disappearance coefficient, $\langle\sin^2(\Delta m^2 L/4E_{\rm{\nue}})\rangle$, which differs between reactor-on and reactor-off ($R^{\rm{\nu}}=R^{\rm{r\nu}}$) data due to the different \nue energy spectrum. This relation between $R_d^{\rm{obs}}$ and $R_d^{\rm{exp}}$ is evaluated for data taken at various reactor conditions, grouping the IBD candidates and the expected \nue in bins of the total baseline-adjusted thermal power ($P^{*}_{\rm{th}}=\sum^{\rm{N_r}}_{i} P^i_{\rm{th}}/L_i^{\rm{2}}$, where $N_r=2$ is the number of cores). A fit of these data points to the model expressed in Eq.~\ref{eq:model} yields the determination of sin$^2(2\theta_{13})$ and $B_d$. As the accidental background rate is known with a precision below 1\%, the fit is performed with accidental-subtracted samples. As a consequence, hereafter $B_d$ refers to all background sources but the accidental one (in short, \emph{cosmogenic} background). The $\eta^{\rm{osc}}$ coefficient is computed, for each $P^*_{\rm{th}}$ bin, by means of simulations as the integration of the normalized antineutrino energy spectrum multiplied by the oscillation effect driven by $\Delta m^2$ (\cite{dm2ee}) and the distance $L$ between the reactor cores and the detectors. The average $\eta_{\rm{FD}}^{\rm{osc}}$ ($\eta_{\rm{ND}}^{\rm{osc}}$) value in the MD reactor-on period is computed to be 0.55 (0.11). The FD coefficient obtained for the SD reactor-on period is slightly larger (0.56) since the relative contributions of \nue from the B1 and B2 cores differ. According to the residual neutrino spectrum discussed in Sec.~\ref{sec:data}, the average $\eta_{\rm{FD}}^{\rm{osc}}$ ($\eta_{\rm{ND}}^{\rm{osc}}$) in the reactor-off period is 0.86 (0.21).

The systematic uncertainties related to the IBD detection and the reactor \nue flux have been described in detail in \cite{dciv}. The same values apply to $R_d^{\rm{exp}}$, although only the normalization uncertainties need to be accounted for. These can be divided in three groups: 1) detection efficiency ($\sigma_{\epsilon}$), 2) reactor-on \nue flux prediction ($\sigma_{\rm{\nu}}$), and 3)  reactor-off \nue flux prediction ($\sigma_{\rm{r}\nu}$). In turn, these errors can be decomposed into their correlated and uncorrelated contributions among the detectors and the reactor cores. The correlated detection efficiency error in the FD and the ND is 0.25\%, while the uncorrelated uncertainties are 0.39\% ($\sigma^{\rm{FD}}_{\epsilon}$) and 0.22\% ($\sigma^{\rm{ND}}_{\epsilon}$), respectively. Since the far detector has the same performance during the SD and MD periods, $\sigma^{\rm{FD}}_{\epsilon}$ is fully correlated in the two data samples. Concerning the reactor-on flux uncertainty, only the thermal power (0.47\%) and fractional fission rates (0.78\%) are considered to be fully uncorrelated among reactors, in a conservative approach. This implies that the total correlated reactor error is 1.41\% (fully dominated by the Bugey4 normalization), while the uncorrelated is 0.91\% ($\sigma_{\rm{\nu}}$) for both B1 and B2 cores. As discussed in \cite{rrm}, the uncertainty on $P_{\rm{th}}$ depends on the thermal power itself. The $P_{\rm{th}}$ error in each $P^*_{\rm{th}}$ bin is computed according to the same procedure. However, as more of 90\% of the reactor-on data are taken at full reactor power (either with 1 or 2 reactors being in operation), the dependence of $\sigma_{\rm{\nu}}$ with $P_{\rm{th}}$ is negligible. In the MD period, $\sigma_{\rm{\nu}}$ is fully correlated among the two detectors, while $\sigma_{\rm{\nu}}$ is conservatively treated as fully uncorrelated between the SD and the MD data. The total correlated normalization error in the expected reactor-on IBD flux $\phi$ in the FD and the ND ($\sigma_{\phi}$) yields 1.43\%, considering the correlated detection and reactor uncertainties. Finally, the uncertainty on the residual \nue in reactor-off periods is treated differently in the SD and MD samples. For the SD sample, $\sigma_{\rm{r}\nu}$ is set to 30\% as estimated in \cite{2off}. For the MD samples, no error is considered as the IBD rate normalization of $R^{\rm{r\nu}}$ is treated as a free parameter in the oscillation fit.

\section{$\theta_{13}$ and background measurements}
\label{sec:results}

The fit of the observed rates for each $P^*_{\rm{th}}$ bin in each detector is based on a standard $\chi^2$ minimization. Apart from the free parameters sin$^2(2\theta_{13})$, $B_d$ and $R^{\rm{r\nu}}$ (only for MD), a set of nuisance parameters $\bar{\alpha}$ are introduced in order to account for the different flux and detection uncertainties. The $\chi^2$ function consists of reactor-on and reactor-off terms, which in turn are divided into individual terms for the FD (SD), FD (MD) and ND detector data samples. In addition, penalty or pull terms are added to constrain each one of the nuisance parameters to the uncertainties quoted in Sec.~\ref{sec:rrm}.

Assuming Gaussian-distributed errors, the reactor-on $\chi^2$ for each bin in the detector sample $d$ is defined as:  

\begin{equation}
\chi^2_{\rm{on},\textit{d}}= \left(\frac{1}{\sigma_{d}^{\rm{stat}}}\right)^2\left(R_{d}^{\rm{obs}}- R_{d}^{\rm{\rm{exp}}}(1+\alpha^{\phi} + \sum_{r=\rm{B1,B2}}\left(w_{d,r}\alpha_{d,r}^{\rm{\nu}}\right)+\alpha^{\epsilon}_{d})-B_{d}\right)^2
\end{equation}

\noindent where $\sigma_{d}^{\rm{stat}}$ is the statistical error and $\alpha^{\phi}$, $\alpha_{d,r}^{\rm{\nu}}$, and $\alpha_d^{\epsilon}$ are the nuisance parameters accounting for the uncertainties $\sigma_{\phi}$, $\sigma_{\rm{\nu}}$ and $\sigma^d_{\epsilon}$, respectively. As $\sigma_{\rm{\nu}}$ is treated as fully uncorrelated between SD and MD, specific $\alpha^{\rm{\nu}}_{d}$ parameters are used in both periods: $\alpha^{\rm{\nu}}_{\rm{FD(SD)},\textit{r}}\neq \alpha^{\rm{\nu}}_{\rm{FD(MD)},\textit{r}} = \alpha^{\rm{\nu}}_{\rm{ND},\textit{r}}\equiv \alpha^{\rm{\nu}}_{r}$. The weights $w_{d,r}$ account for the relative fraction of antineutrinos generated in the reactor $r$ and detected in the detector $d$. These values are computed according to the Monte Carlo simulation considering the reactor powers and the baselines. Due to the low statistics in the reactor-off periods, specially in the SD one, the uncertainty in the sample of selected events is considered to be Poisson-distributed. The reactor-off $\chi^2$ is then defined as binned Poisson likelihood following a $\chi^2$ distribution:

\begin{equation}
\label{eq:chi2off}
\chi^2_{\rm{off},\textit{d}} = 2 \Big( N_d^{\rm{obs}}\textrm{ln} \frac{N_d^{\rm{obs}}}{C_d+N_d^{\rm{exp}}[1+\alpha^{\epsilon}_{d}+\alpha^{\rm{r\nu}}_d]}+C_{d}+N_d^{\rm{exp}}[1+\alpha^{\epsilon}_{d}+\alpha^{\rm{r\nu}}_d] - N_d^{\rm{obs}}\Big)
\end{equation}

\noindent where $N^{\rm{obs}}$ is the number of observed IBD candidates, $C$ is the number of cosmogenic background events ($C=B\times T$, being $T$ the reactor-off live time), $N^{\rm{exp}}$ is the expected number of antineutrinos ($N^{\rm{exp}}=R^{\rm{exp}}\times T$),  and $\alpha^{\rm{r\nu}}_d$ is the parameter accounting for the error on the residual \nue expectation. While in the SD reactor-off data this parameter is constrained by $\sigma_{\rm{r\nu}}$, in the MD reactor-off it is left free, but correlated between the FD and the ND according to the ratio of reactor-averaged baselines $L_d$: $\alpha^{\rm{r\nu}}_{\rm{ND}}=L_{\rm{FD}}^2/L_{\rm{ND}}^2 \times \alpha^{\rm{r\nu}}_{\rm{FD}}$. Finally, The last term of the $\chi^2$ incorporates Gaussian pulls for the $\bar{\alpha}$ parameters according to their associated uncertainties:

\begin{equation}
\chi^{2}_{\rm{pull}} = \left(\frac{\alpha^{\phi}}{\sigma_{\phi}}\right)^2 + \left(\frac{\alpha^{\epsilon}_{\rm{FD}}}{\sigma^{\rm{FD}}_{\epsilon}}\right)^2 +\left(\frac{\alpha^{\epsilon}_{\rm{ND}}}{\sigma^{\rm{ND}}_{\epsilon}}\right)^2 + \sum_{r}^{N_r}\left(\frac{\alpha^{\rm{\nu}}_{r}}{\sigma_{\rm{\nu}}}\right)^2+ \sum_{r}^{N_r}\left(\frac{\alpha^{\rm{\nu}}_{\rm{FD(SD)},r}}{\sigma_{\rm{\nu}}}\right)^2 + \left(\frac{\alpha^{\rm{r\nu}}_{\rm{FD(MD)}}}{\sigma_{\rm{r\nu}}}\right)^2 
\end{equation}

The number of reactor-on bins considered for each data sample ($N_{\rm{b}}$) has been set according to the available statistics. As done in \cite{rrm}, the SD data is divided in 6 $P^*_{\rm{th}}$ bins, while the MD data in 4 bins for both detectors. The overall $\chi^2$ function used for the fit can then be expressed as:

\begin{equation}
\label{eq:rrm}
\chi^2 = \sum_d \left( \sum_i^{N_{\rm{b}}} \chi^2_{\rm{on},\textit{d},\textit{i}} + \chi^2_{\rm{off},\textit{d}}\right) +\chi^2_{\rm{pull}}
\end{equation}

\begin{figure}
  \begin{center}
    \includegraphics[scale=0.37]{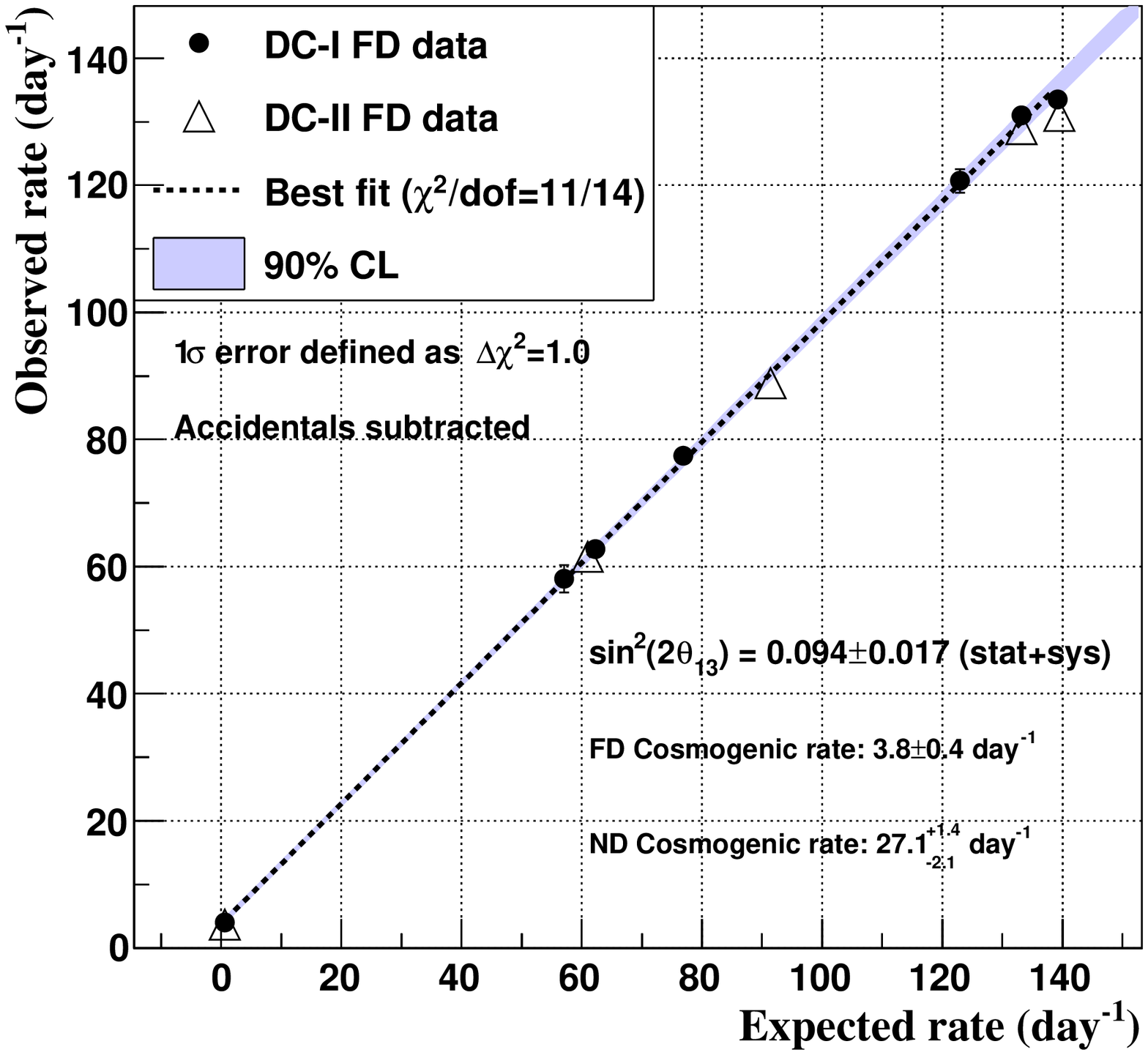}
    \includegraphics[scale=0.37]{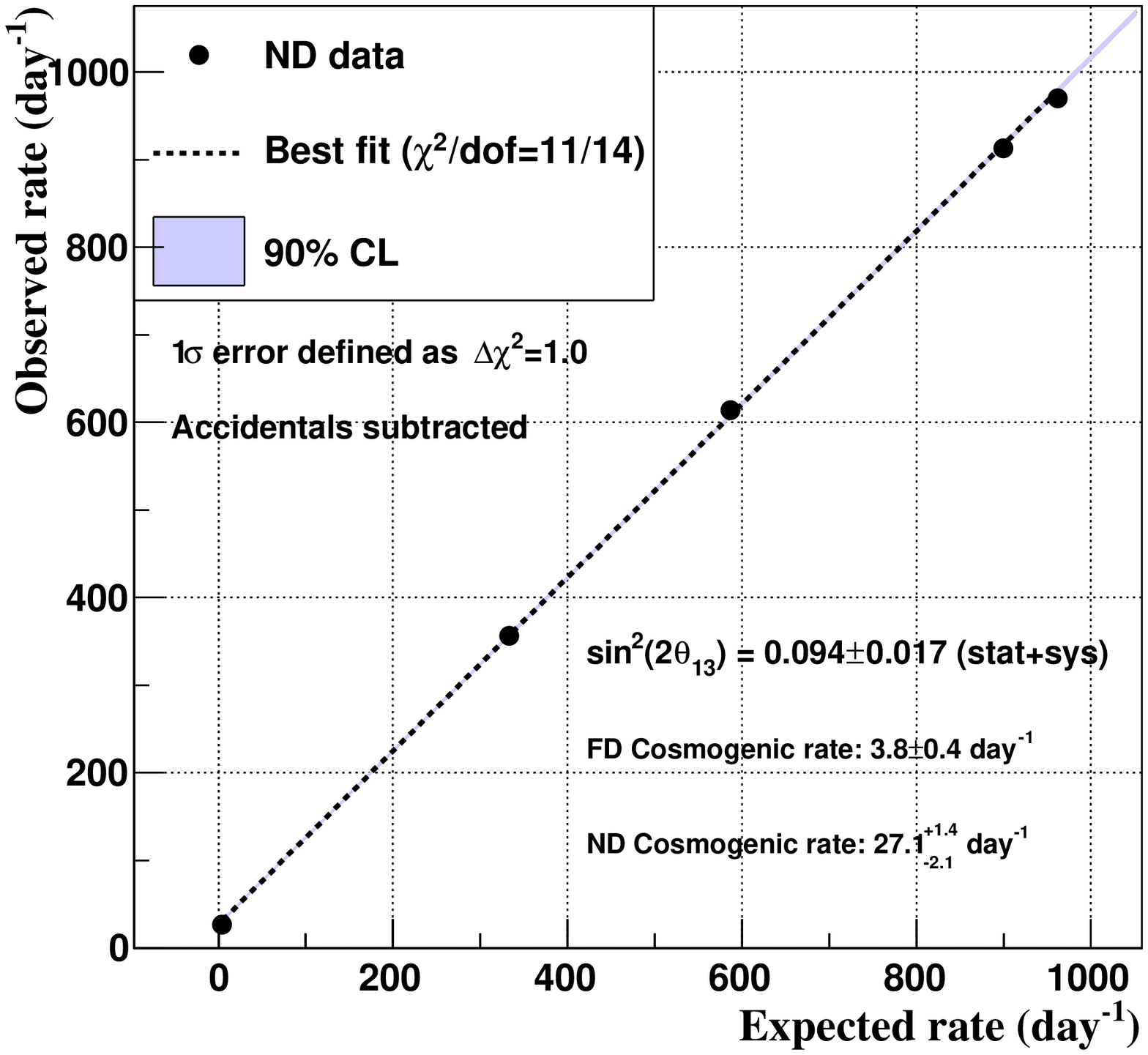}
    \includegraphics[scale=0.37]{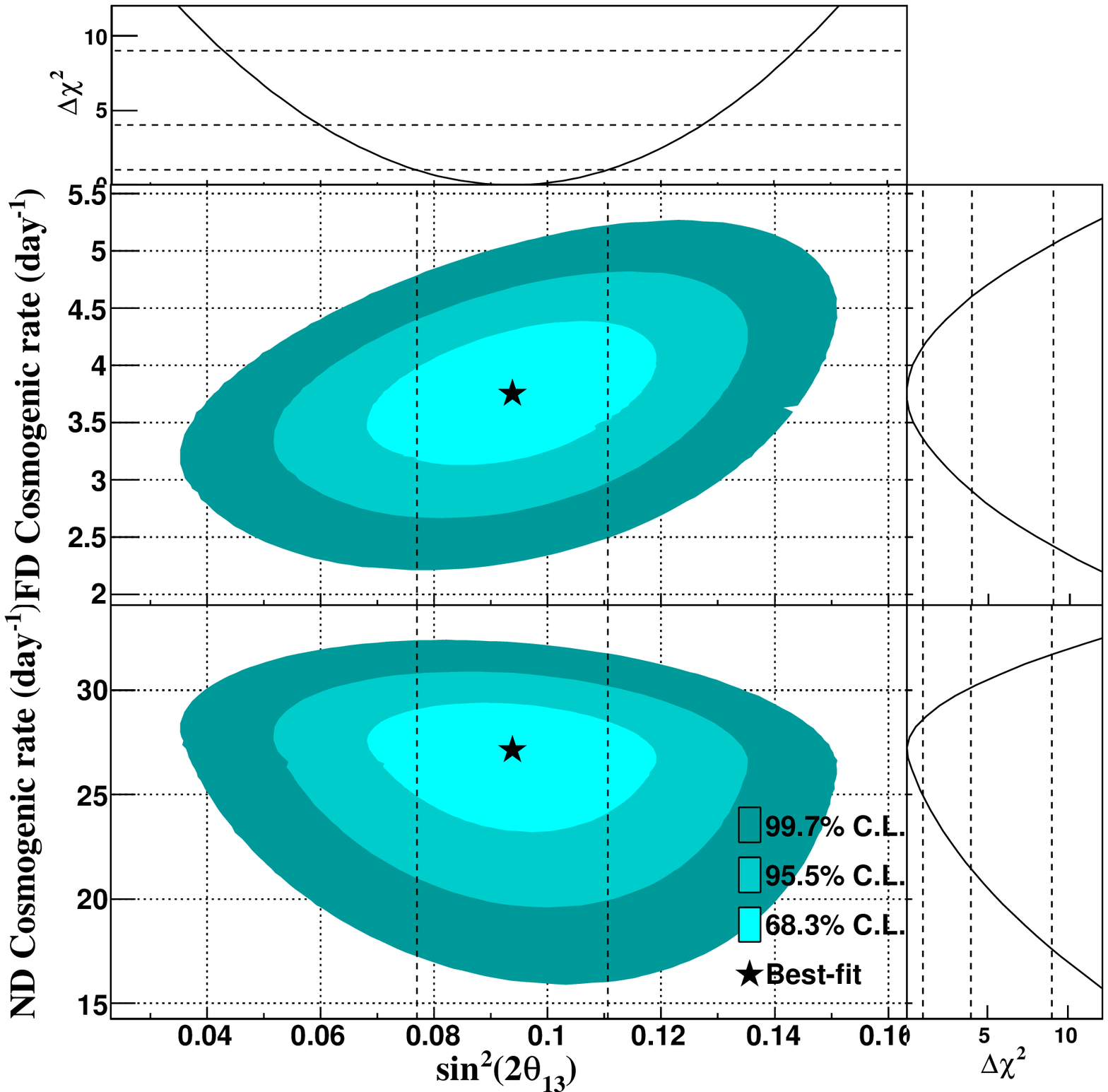}
    \caption{ RRM (sin$^2(2\theta_{13})$, $B_{\rm{FD}}$, $B_{\rm{ND}}$) fit results. The upper plots show the observed rates versus the expected rates in the FD (left) and the ND (right), superimposed to the best-fit model (dashed line). The statistical errors, not visible, are at the level of $\sim1\%$ ($\sim0.4\%$) in the FD (ND). The bottom plot shows the 68.4\%, 95.5\% and 99.7\% C.L. regions for the three parameters and the one-dimensional projections of $\Delta\chi^{2}$.}
    \label{fig:th13bgindep}
  \end{center}
\end{figure}

The results of the (sin$^2(2\theta_{13})$, $B_{\rm{FD}}$, $B_{\rm{ND}}$) fit are shown in Fig.~\ref{fig:th13bgindep}, in terms of the observed versus the expected rate, and of the 68.4\%, 95.5\% and 99.7\% confidence level (C.L.) regions. The RRM yields best-fit values of sin$^2(2\theta_{13})=0.094\pm0.017$, $B_{\rm{FD}}=3.75\pm0.39~\rm{day}^{-1}$ and $B_{\rm{ND}}=27.1^{+1.4}_{-2.1}~\rm{day}^{-1}$, with $\chi^{2}/dof=11.0/14$. The background-independent determination of $\theta_{13}$ is consistent with all previous results form Double Chooz, being the precision competitive with the one achieved by the R+S analysis in \cite{dciv}. The values of the total cosmogenic backgrounds in the FD and ND are also consistent with the sum of the $^9$Li and fast-neutron background expectations quoted in Tab.~\ref{tab:bg}, respectively, with similar associated errors. The best-fit for the residual neutrinos is brought to the physical limit of 0, with $R^{\rm{r\nu}}_{\rm{FD}}<0.64~\rm{day}^{-1}$ and $R^{\rm{r\nu}}_{\rm{ND}}<4.2~\rm{day}^{-1}$ at 90\% C.L.

\begin{figure}
  \begin{center}
    \includegraphics[scale=0.37]{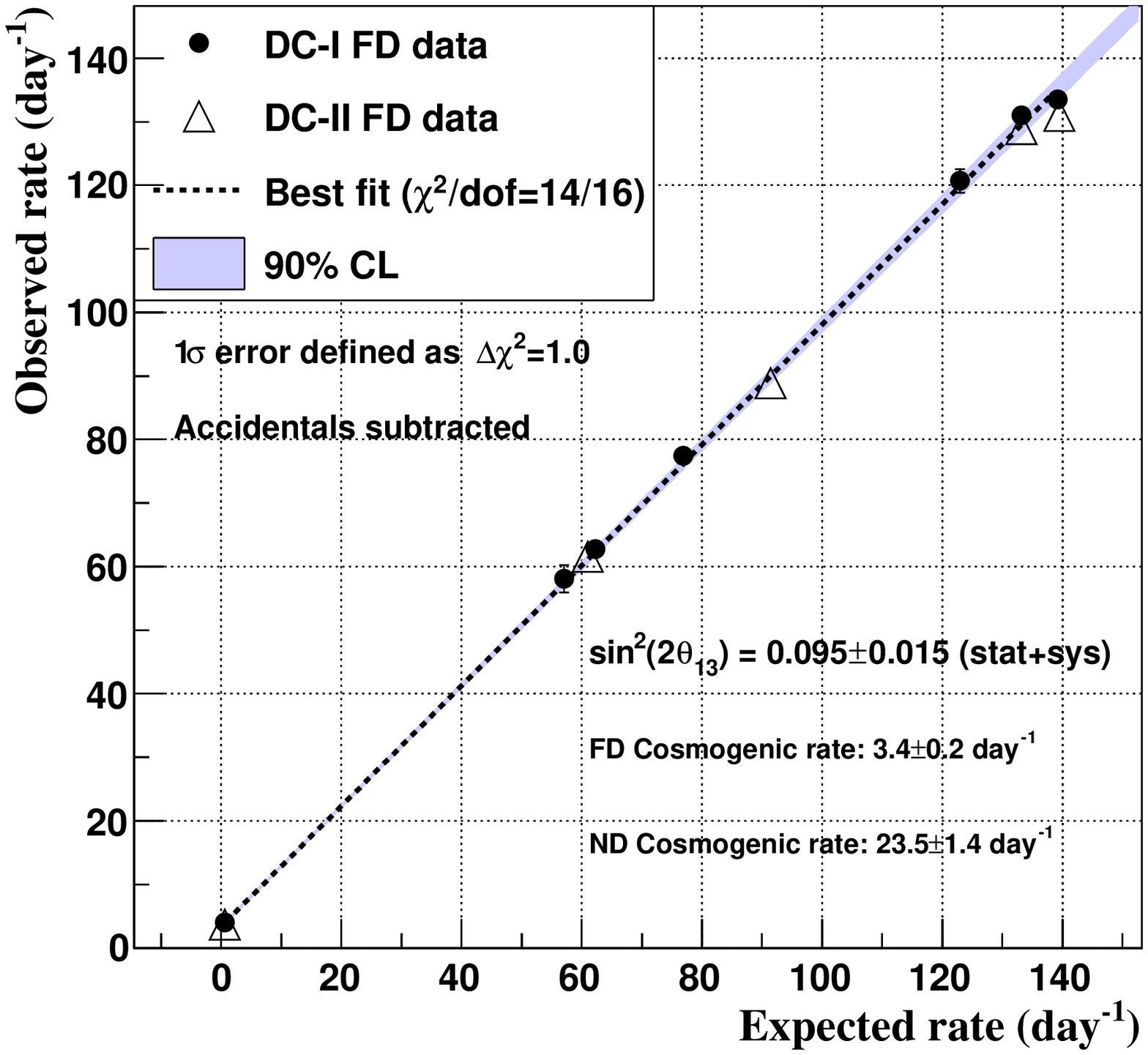}
    \includegraphics[scale=0.37]{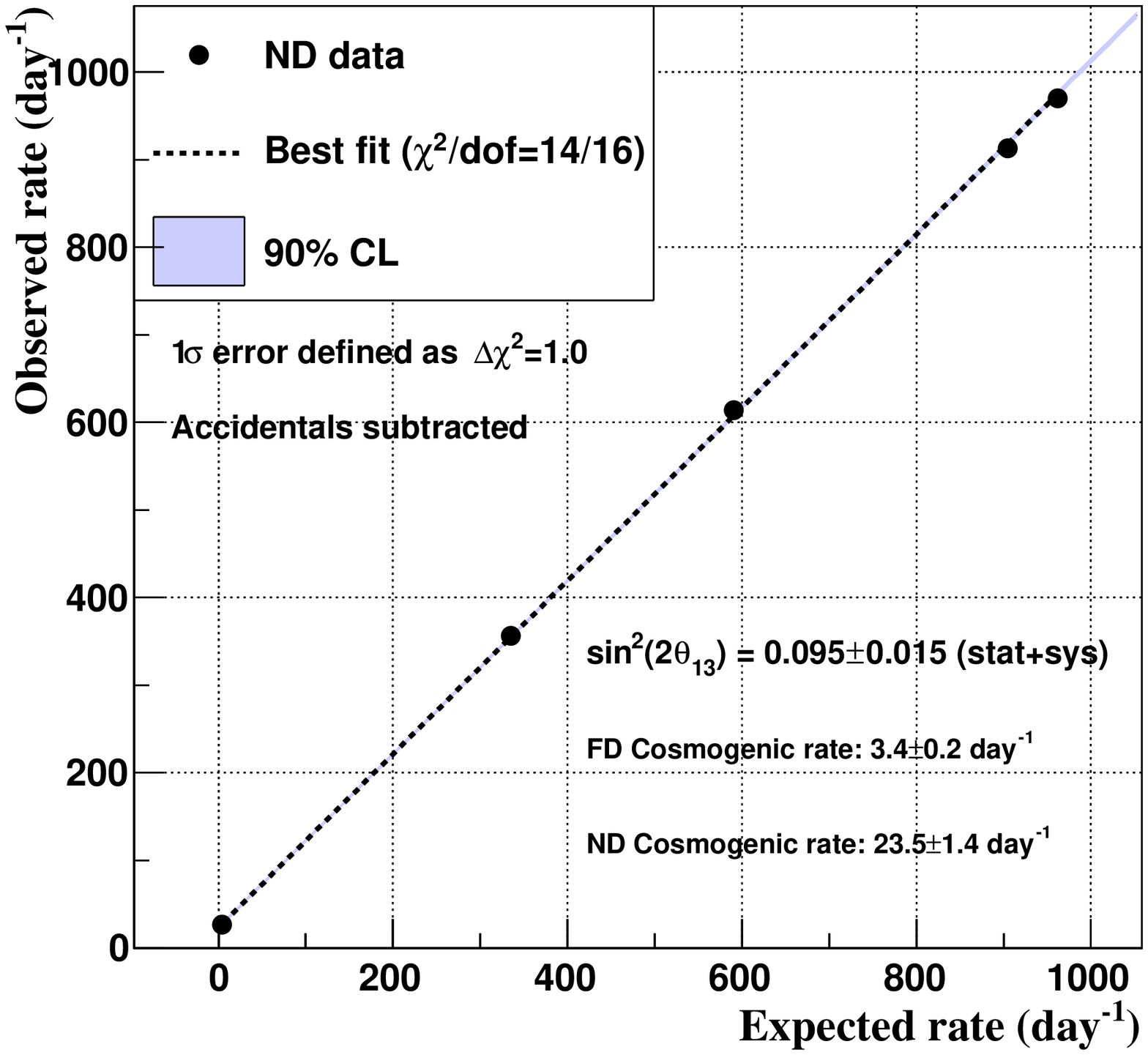}
    \caption{ RRM sin$^2(2\theta_{13})$ fit results. The observed rates versus the expected rates in the FD (left) and the ND (right) are shown superimposed to the best-fit model (dashed line). The statistical errors, not visible, are at the level of $\sim1\%$ ($\sim0.4\%$) in the FD (ND). The parameters $B_{FD}$ and $B_{ND}$ are constrained by the estimates obtained from reactor-on data.}
    \label{fig:th13bg}
  \end{center}
\end{figure}

The precision on $\theta_{13}$ can be improved by introducing the constraint of the cosmogenic background estimates into the fit. The background constraint is added to the $\chi^2$ function as two extra Gaussian priors:

\begin{equation}
\chi^2_{\rm{BG}} =  \left(\frac{B^{\rm{exp}}_{\rm{FD}}-B_{\rm{FD}}}{\sigma_{\rm{B}}^{\rm{ND}}}\right)^2 +\left(\frac{B^{\rm{exp}}_{\rm{ND}}-B_{\rm{ND}}}{\sigma_{\rm{B}}^{\rm{ND}}}\right)^2  
\end{equation}

\noindent where $B^{\rm{exp}}$ and $\sigma_{\rm{B}}$ stand for the central value and uncertainty of the background expectations. These are built considering the fast-neutron determination from reactor-on data (Tab.~\ref{tab:bg}) and the combination of the $^9$Li determinations with reactor-on and reactor-off data: $B^{\rm{exp}}_{\rm{FD}}=3.33\pm0.29~\rm{day}^{-1}$ and $B^{\rm{exp}}_{\rm{ND}}=22.57\pm1.55~\rm{day}^{-1}$. The results of the corresponding sin$^2(2\theta_{13})$ fit are presented in Fig.~\ref{fig:th13bg}. The fit yields a best-fit value of sin$^2(2\theta_{13})=0.095\pm0.015$, with $\chi^{2}/dof=13.5/16$. As expected due to the consistency between the background estimates and the reactor-off data, the central value is not significantly modified with respect to the background-independent $\theta_{13}$ result. However, the combination of the background model and the reactor-off information allows for a more precise determination of the residual neutrinos, yielding now $R^{\rm{r\nu}}_{\rm{FD}}=0.48\pm0.28~\rm{day}^{-1}$ and $R^{\rm{r\nu}}_{\rm{ND}}=3.18\pm1.85~\rm{day}^{-1}$. According to these non-vanishing values, the best-fit parameters of $B_{\rm{FD}}$ and $B_{\rm{ND}}$ ($3.37\pm0.24~\rm{day}^{-1}$ and $23.49\pm1.40~\rm{day}^{-1}$, respectively) are slightly reduced with respect to the background-model-independent fit results.

Although fully consistent, the best fit of $B_{\rm{ND}}$ is found to be $\sim$7\% higher than the one obtained in \cite{dciv} ($21.86\pm1.33$~$\rm{day}^{-1}$ in the 1.0-8.5 MeV energy range). This difference is mostly driven by the background constraint and the use of the new MD reactor-off data, not considered in the R+S analysis yet. In Tab.~\ref{tab:bfbg}, a summary of the background expectations and the best-fit values is presented. Due to the anti-correlation between sin$^2(2\theta_{13})$ and the background in the ND (visible in bottom plot of Fig.~\ref{fig:th13bgindep}), the larger best-fit value of $B_{\rm{ND}}$ obtained in the current RRM analysis pulls down the central value of $\theta_{13}$ with respect to the R+S result (sin$^2(2\theta_{13})=0.105\pm0.014$). 



\begin{table}[!htb]
\caption{\label{tab:bfbg} Cosmogenic background expectations and best-fit values in the FD and the ND. For comparison purposes, the last row shows the best-fit values obtained in the R+S analysis \cite{dciv}, restricted to the 1.0-8.5 MeV energy window.}
\begin{center}
\begin{tabular}{ccc}
  \hline
  Background (day$^{-1}$)    &   FD              &  ND \\ \hline
  Expectation               &   3.33$\pm$0.29   &  22.57$\pm$1.55\\
  RRM Best-fit              &   3.37$\pm$0.24   &  23.49$\pm$1.40\\
  \hline
  R+S Best-fit              &   3.43$\pm$0.25   &  21.86$\pm$1.33\\
  \hline
\end{tabular}
\end{center}
\end{table}

Finally, in order to provide a measurement of the IBD rate normalization, the parameter $\alpha^{\phi}$ can be left free in the fit by removing the corresponding pull term in the $\chi^{2}$. Since the correlated detection efficiency is known with a negligible uncertainty (0.25\%), this parameter provides effectively the relative normalization with respect to the central value of the reactor-on flux simulation. This central value is defined by the mean cross-section per fission (see \cite{dciv} for details), $\langle \sigma_{\rm{f}} \rangle$, measured by Bugey-4. Once corrected for the specific averaged fuel compositions of the Double Chooz reactor cores, it is computed to be $(5.75\pm0.08)\times10^{-43}\rm{cm}^{2}/\rm{fission}$. The best-fit value of  $\alpha^{\phi}$ yields $0.04\pm0.86\%$, thus being fully consistent with the expectation but reducing the 1.43\% uncertainty on the IBD rate normalization $\sigma_{\phi}$. According to this result, the best-fit of sin$^2(2\theta_{13})$ is not significantly modified.

\section{Summary and conclusions}
\label{sec:conclusions}

The simple experimental setup of Double Chooz, consisting of only two detectors and two reactors, allows for a simultaneous determination of $\theta_{13}$ and the total background rates. The RRM analysis relies on the rate of observed \nue interactions in data samples collected at different total reactor powers. The comparison of such rates with the null-oscillation Monte Carlo expectations provides a background-independent measurement of sin$^2(2\theta_{13})$, as well as inclusive background rates in the far and near detectors which do no depend of any a priori assumptions on the individual background sources. This approach is intrinsically different from the usual R+S $\theta_{13}$ oscillation analyses implemented in reactor experiments, which are based on background models considering a number of background sources and estimating the corresponding rates and energy spectra from reactor-on data.

In this work, a multi-detector RRM analysis is implemented for the first time. As in R+S analyses, the relative comparison of the rates observed at different baselines leads to a major reduction of the involved systematic uncertainties. In particular, the correlated detection and reactor \nue flux errors cancel out, while the uncorrelated flux uncertainty is significantly suppressed. Apart from boosting the precision in the $\theta_{13}$ measurement, the multi-detector measurement of the IBD interactions allows for a determination of the observed IBD rate normalization. The current oscillation analysis also uses for the first time reactor-off data samples for both the FD and ND, offering a powerful handle to constrain the backgrounds. Among the $\theta_{13}$ reactor experiments, Double Chooz is the only one with available reactor-off samples, thus offering a unique cross-check of the background models.

The RRM oscillation fit relies on the minimization of a $\chi^{2}$ function consisting of reactor-on and reactor-off terms, as well as penalty terms constraining the nuisance parameters accounting for the systematic uncertainties to their estimated values. The errors considered in the fit are those impacting the expected IBD rates, namely, the detection efficiency and the reactor flux normalization uncertainties. The (sin$^2(2\theta_{13})$, $B_{\rm{FD}}$, $B_{\rm{ND}}$) fit yields a background-independent value of $\theta_{13}$ which is consistent with previous Double Chooz results: sin$^2(2\theta_{13})=0.094\pm0.017$. The precision achieved by the RRM analysis is competitive with that one obtained in the R+S fit presented in \cite{dciv} (sin$^2(2\theta_{13})=0.105\pm0.014$), relying on a reactor-on background model. The best fit values of the total cosmogenic background rates in the FD and ND, $B_{\rm{FD}}=3.75\pm0.39~\rm{day}^{-1}$ and $B_{\rm{ND}}=27.1^{+1.4}_{-2.1}~\rm{day}^{-1}$, are also consistent with the background estimates. Thus, these expectations can be added to the RRM in order to improve the precision of the oscillation results: sin$^2(2\theta_{13})=0.095\pm0.015$. The limited reduction on the error is due to the dominant role of the detection and reactor flux systematic uncertainties. The compatibility of the background-model-dependent R+S and the RRM results, as well as the consistency between the reactor-off data and the background models, confirms the robustness of the Double Chooz oscillation analyses. Beyond the $\theta_{13}$ result, the RRM fit is used to measure the observed \nue rate normalization. The best-fit value yields a $0.04\pm0.86\%$ deviation with respect to the flux normalization predicted by Bugey-4, thus being fully consistent.

\acknowledgments
We thank the company {\it Electricit\'e de France} (EDF); 
the European fund FEDER; 
the R\'egion de Champagne-Ardenne; 
the D\'epartement des Ardennes;
and the Communaut\'e de Communes Ardenne Rives de Meuse.
We acknowledge the support of 
the CEA, CNRS/IN2P3, the computer centre CC-IN2P3
and 
LabEx UnivEarthS in France;
the Max Planck Gesellschaft, 
the Deutsche Forschungsgemeinschaft DFG, 
the Transregional Collaborative Research Center TR27, 
the excellence cluster ``Origin and Structure of the Universe''
and 
the Maier-Leibnitz-Laboratorium Garching in Germany;
the Ministry of Education, Culture, Sports, Science and Technology of Japan (MEXT) 
and
the Japan Society for the Promotion of Science (JSPS) in Japan;
the Ministerio de Econom\'ia, Industria y Competitividad (SEIDI-MINECO) under grants FPA2016-77347-C2-1-P and MdM-2015-0509 in Spain;
the Department of Energy and the National Science Foundation 
and
Department of Energy in the United States;
the Russian Academy of Science, 
the Kurchatov Institute 
and 
the Russian Foundation for Basic Research (RFBR) in Russia;
the Brazilian Ministry of Science, Technology and Innovation (MCTI), 
the Financiadora de Estudos e Projetos (FINEP), 
the Conselho Nacional de Desenvolvimento Cient\'ifico e Tecnol\'ogico (CNPq), 
the S\~ao Paulo Research Foundation (FAPESP)
and 
the Brazilian Network for High Energy Physics (RENAFAE) in Brazil.

\bibliographystyle{JHEP}
\bibliography{biblio}

\end{document}